\documentstyle[sprocl]{article}

\def\Journal#1#2#3#4{{#1} {\bf #2}, #3 (#4)}

\def\NPB{{\em Nucl. Phys.} B}
\def\PLB{{\em Phys. Lett.}  B}
\def\PRL{\em Phys. Rev. Lett.}
\def\PRD{{\em Phys. Rev.} D}
\def\PR{\em Phys. Rept.} 

\def\ZPB{{\em Z. Phys.} B} 
\def\APPB{{\em Acta Phys. Polon.} B}
\def\IJMPA{{\em Int.J.Mod.Phys.} A}
\def\AP{\em Ann. Physik}

\def\be{\begin{equation}}
\def\ee{\end{equation}}
\def\bea{\begin{eqnarray}}
\def\eea{\end{eqnarray}}

\begin{document}

\title{FLOW EQUATIONS IN THE LIGHT-FRONT QCD} 

\author{E. GUBANKOVA}

\address{Department of Physics, North Carolina State University,
Raleigh,\\ NC 27695-8202, USA\\E-mail: egoubank@unity.ncsu.edu}


\maketitle\abstracts{
Flow equations method of continuous unitary transformations
is used to eliminate the minimal quark-gluon interaction 
in the light-front quantized QCD Hamiltonian.
The coupled differential equations in the two lowest Fock sectors  
correspond to the renormalization
of the light-front gluon mass and the generation 
of effective quark-antiquark interaction.
The influence of the renormalization of the gluon
effective mass on the elimination of the quark-gluon coupling
and the induced quark-antiquark interaction
is taken into account. Namely,  
the original gauge field coupling can be completely
eliminated, even when the states connected by this
interaction are degenerate. Furthermore,
even in the case where effective interaction,
obtained within perturbative schemes
(bound state perturbation theory 
or perturbative similarity approach),
is not defined, we obtain more singular behavior
$1/q^4$ at small gluon momenta.
This is due to asymptotic behavior of the effective
gluon mass at small cutoffs. By discussing
the consequences of this asymptotic behavior,
it seems that our approach is superior
to perturbation theory and to perturbative similarity
approach.}

\section{Introduction}

The perturbative aspects of non-abelian gauge theories
were underestood many years ago, and the perturbative calculations 
provided convincing proof that QCD is the theory of strong interactions. 
However nonperturbative QCD phenomena have been difficult  
to analyze mainly because calculational techniques are still lacking,
even though the qualitative features have been more or less
understood.

In particular, it is widely believed that pure Yang-Mills theory,
with no dynamical quarks, posseses the features
like asymptotic freedom, mass generation through the transmutation
of dimensions, and confinement: linear rising potential between static 
(probe) quarks. Adding dynamical quarks chiral symmetry 
is broken spontaneously. 
The ultimate aim of this study is to understand these 
nonperturbative mechanisms in a Hamiltonian framework,
solving flow equations for canonical QCD Hamiltonian  
in the few lowest sectors selfconsistently for dynamical 
qluons and quarks, and their effective interactions.
In this work dynamics of quarks has been excluded to disentangle
the complexity of chiral symmetry breaking.

In the past few years there were several studies  
addressing the issue of confinement and generation of mass gap
in the framework of the Schr{\"o}dinger picture
\cite{DiakonovZarembo}, \cite{Karabali}. In the relatively recent works 
\cite{DiakonovZarembo} one have been using the special ansatz
for a vacuum wave functional suggested by Kogan and Kovner,
and integrating over all possible gauge configurations.
To mention a few early works, see refs. \cite{Schr"odinger}. 
The calculational technique in this approach is still rather complicated,
and allows to solve a field theory problem only in $1+1$ dimensions 
\cite{Karabali}, but treats only ground states in 
$3+1$ dimensions \cite{DiakonovZarembo}. 

In alternative studies of nonperturbative problems 
in Hamiltonian framework
one considers directly the QCD Hamiltonian quantized 
in a special gauge,
in particular the light-front gauge, $A^+=0$
\cite{BrodskyPauliPinsky}. 
There is a belief that the light-front gauge may be the most 
suitable frame to study the nonperturbative QCD 
\cite{BrodskyPauliPinsky}.
Previous studies of confinement and bound states in the light-front frame
have been done using the methods 
of similarity renormalization \cite{BrisudovaPerry}
and transverse lattice \cite{BvdSandeDalley},
and are based on the fact that the light-front QCD in $3+1$ already 
has a confining interaction in the form of the instantaneous 
four-fermion interaction, $1/q^{+~2}$, 
which is the confining interaction in $1+1$
along the light-front direction, $x^-$ . However, since instantaneous 
interaction does not provide confinement mechanism for quarks and gluons
in $3+1$ dimensions, the task of maintenance rotational symmetry becomes 
difficult to achieve when trying to extend 
light-front confinement to $3+1$.

Some time ago Wilson proposed a formalism to construct a confining
light-front quark-gluon Hamiltonian for light-front 
$QCD_{3+1}$ \cite{Wilson}. 
Wilson suggested that a starting point for analyzing the full QCD 
with confinement in light-front coordinates is the 
light-front infrared divergences. 
Based on light-front power counting, the counterterms for the 
light-front infrared
divergences can involve the color charge densities and involve unknown 
nonlocal behavior in transverse direction that become a possible source
for transverse confinement. However, the analysis was not complete
and a scheme for practical calculation has not been developed.

A subtle point in the light-front field theory is that 
the light-front vacuum is just empty space. 
Therefore it seems a problem how confinement can occur 
in the light-front frame, and what quantity sets up a scale for a dynamical
mass gap and the string tension. The infrared longitudinal cutoff  
properties of light-front theory suggest a fundamental role for 
the light-front counterterms, solving this paradox.
Namely, the longitudinal infrared cutoff in light-front dynamics
makes it impossible to create particles from a bare vacuum 
by a translationally invariant Hamiltonian and in addition
the number of constituents in a given eigenstate is limited. 
When introducing infrared cutoff physics below the cutoff is missed, 
and to restore it one may introduce appropriate counterterms.
Light-front counterterms to the longitudinal infrared cutoff
dependence provide a nonzero amplitude of particle creation,  
and become therefore a possible alternative source for features 
normally associted in standard equal-time dynamics with 
a nontrivial vacuum structure, including confinement and 
spontaneous symmetry breaking. Note, that small light-front $x$
correspond to high light-front energies. Therefore in order
to remove small $x$, one should use renormalization group.
 
To be more specific we adopt the following model 
suggested by Susskind and Burkardt in the context 
of chiral symmetry breaking 
in the light-front frame \cite{SusskindBurkardt}.  
In the parton model one pictures a fast moving hadron 
as being some collection of constituents with relatively large momentum,
such that when one boosts the system, doubles its momentum,
all these partons double their momenta and so forth.
Therefore one can formulate an effective field theory on the axis
of the light-front momentum $x$ (or on rapidity axis, 
which is logarithm of $x$). 
Partons that form a hadron are at positive, finite $x$ and according
to Feynman and Bjorken fill the $x$-axis in a way which gets 
denser and denser as one goes to smaller $x$; and the vacuum is
at $x=0$. The fundamental property of light-front Hamiltonians, 
that under a rescaling of the light-front momentum,
$x\rightarrow \lambda x$, the light-front Hamiltonian scales like
$H\rightarrow H/\lambda$, can be interpreted as a dilatation symmetry
along the $x$-axis, if one thinks of the $x$-axis as a spacial axis.
This symmetry holds on a classical level and it is broken
on a quantum level by anomali.
As one approaches small $x$, interaction between partons gets stronger,
contributing divergent matrix elements. A natural cutoff
is provided by $\delta x=\varepsilon^{+} x$, 
a minimal spacing between constituents,
which plays the role of UV-regulator. 
As long as $\delta x$ (or $\varepsilon^{+}$) is finite, 
i.e. as long as the density of partons
on the $x$-axis is not infinite, one obtains finite matrix elements.
Cutoff $\delta x$ breaks the dilatation symmetry along the $x$-axis
and gauge symmetry, and sets up an energy scale in effective 
light-front field theory formulated on $x$-axis. 
In terms of effective theory a generated mass gap defines 
a strength of effective interactions between quarks,
in this case string tension in quark-antiquark potential.
Formation of the $q\bar{q}$ bound state through breaking 
an internal symmetry can be viewed analogously 
to the creation of Cooper pairs in superconducter.

The dilatation symmetry reflects some underlying scale
invariance of the light-front field theory formulated on $x$-axis. 
Introducing the cutoff, breaks this symmetry. 
Because physics should remain independent from the cutoff, 
one must be looking for a fixed point of 
the renormalization group.
Therefore the right tool for studing 
such a system is the renormalization group, which is provided
by the method of flow equations for Hamiltonians 
\cite{Wegner}.
  
Incorporating effects from small $x$ 
into an effective light-front Hamiltonian is equivalent to integrating
out high light-front energy modes in asymtotically free domain. 
In terms of renormalization group, regulating small $x$ introduces 
a mass gap, which together with asymptotic freedom leads
to a renormalization group invariant scale and dimensional transmutation  
along $x$-axis. Mass gap and a possible, more singular 
than a Coulomb, confining potential between quark and antiquark
are direct consequences of dimensional transmutation in the effective
light-front field theory, namely the light-front QCD, 
formulated on the light-front $x$-axis.

In the main part of the paper, basing on the QCD Hamiltonian
in the light-front gauge, flow equations for an effective
gluon mass and effective quark-antiquark interaction
are formulated in the light-front frame and 
solved selfconsistently within the leading iteration. 
By discussing solutions of these equations in concluding section,
it seems that flow equation method is superior
to perturbation theory and perturbative similarity approach.

\section{Flow equations in the light-front QCD}

We apply flow equations to the light-front QCD Hamiltonian
in order to eliminate the minimal quark-gluon interaction, namely to
decouple matter and gauge degrees of freedom in the leading order. 
In the two lowest Fock sectors of 
the effective QCD Hamiltonian 
we obtain coupled differential equations 
which correspond to renormalization of the light-front gluon mass
(renormalization of the quark mass is not considered)
and generation of an effective quark-antiquark interaction.
These flow equations are solved selfconsistently 
in the sense that the influence of the gluon mass renormalization 
on the elimination of the quark-gluon coupling and
the induced quark-antiquark interaction is taken into account.
As a result, the original gauge field coupling is completely 
eliminated, even when the states connected by this interaction 
are degenerate. Furthermore, in the degenerate case where 
effective $q\bar{q}$ interaction obtained within perturbation
theory is not defined, we obtain more singular 
$1/q^4$ behavior at small gluon momenta.

\subsection{Gluon gap equation}

Coupled system of the light-front equations for 
the effective quark and gluon masses as functions 
of a cut-off $\lambda$
was derived first by Glazek \cite{Glazek}. We decouple 
this system of equations by neglecting the cut-off dependence 
of the quark mass, i.e. $m(\lambda)=m$ with $m$ current
quark mass. 
The light-front gluon gap equation reads
\begin{eqnarray}
\frac{d\mu^2(\lambda)}{d\lambda} =
&-& 2T_fN_f g^2\int_0^1\frac{dx}{x(1-x)}
\int_0^{\infty}\frac{d^2k_{\perp}}{16\pi^3}
\frac{1}{Q^2_2(\lambda)}
\frac{df^2(Q^2_2(\lambda);\lambda)}{d\lambda}
\nonumber\\
&\times&
\left(
\frac{k_{\perp}^2+m^2}{x(1-x)}
-2k_{\perp}^2\right)
\nonumber\\
&-& 2C_a g^2\int_0^1\frac{dx}{x(1-x)}
\int_0^{\infty}\frac{d^2k_{\perp}}{16\pi^3}
\frac{1}{Q_1^2(\lambda)}
\frac{df^2(Q_1^2(\lambda);\lambda)}{d\lambda}
\nonumber\\
&\times&
\left(
k_{\perp}^2(1+\frac{1}{x^2}+\frac{1}{(1-x)^2})
\right)
\,,\label{eq:2.1}\end{eqnarray}
with
\begin{eqnarray}
Q_1^2(\lambda) &=& \frac{k_{\perp}^2+\mu^2(\lambda)}{x(1-x)}
-\mu^2(\lambda)
~,~
Q_2^2(\lambda) =
\frac{k_{\perp}^2+m^2}{x(1-x)}
-\mu^2(\lambda)
\,,\label{eq:2.2}\end{eqnarray}
where in the integral kernel gluon couples to the quark-antiquark pairs
and pairs of gluons with the bare strong coupling $g$;
$(x,k_{\perp})$ is the light-front momentum in the loops.
Group factors for $SU(N_c)$ are
$T_f\delta_{ab}={\rm Tr}(T^aT^b)=\frac{1}{2}\delta_{ab}$
and the adjoint Casimir 
$C_a\delta_{ab}=f^{acd}f^{bcd}=N_c\delta_{ab}$,
$N_c$ is the number of colors (i.e., $N_c=3$).

Solution of this gap equation defines an effective
gluon mass at zero gluon momentum and
within the leading iteration
reads (for details see \cite{GubankovaJiCotanch})
\begin{eqnarray}
\mu^2(\lambda) &=& 
\mu_0^2+\delta\mu_{PT}^2(\lambda)
+\delta\mu^2(\lambda,\lambda_0)
\,.\label{eq:2.3}\end{eqnarray}
where $\mu_0$ is the 'physical' mass parameter, 
which fixes the resulting effective gluon mass 
at the scale $\lambda_0\rightarrow 0$ as 
$\mu_{ren}^2(\lambda=\lambda_0)=\mu_0^2$ (see below);
the perturbative term
\begin{eqnarray}
\delta\mu_{PT}^2(\lambda)=-\frac{g^2}{4\pi^2}
\lambda^2 \left(
C_a (\ln\frac{u^2}{\mu_0^2}-\frac{11}{12})
+T_fN_f\frac{1}{3}\right)
\,.\label{eq:2.4}\end{eqnarray}
reproduces the result of the light-front perturbation theory
\cite{ZhangHarindranath}.
Renormalizing the effective Hamiltonian 
through the second oder in coupling $O(g^2)$,
the perturbative mass correction is absorbed by 
the mass counterterm,
$m_{CT}^2(\Lambda_{UV})=-\delta\mu_{PT}^2(\Lambda_{UV})$
with $\Lambda_{UV}\rightarrow\infty$,
and the renormalized effective gluon mass reads 
$\mu^2_{ren}(\lambda)=\mu^2(\lambda)+m_{CT}^2(\lambda)$
for $\lambda=\Lambda_{UV}\rightarrow\infty$.
Though the perturbative renormalization is applied at large
cut-off scales, $\Lambda_{UV}$, we assume that the leading cutoff
dependence is absorbed by the mass counterterm for all $\lambda$.
Therefore the resulting effective mass, renormalized in the second 
order, is given 
\begin{eqnarray}
\mu_{ren}^2(\lambda) &=& \mu_0^2+\delta\mu^2(\lambda,\lambda_0)
= \mu_0^2+\sigma(\mu_0,u)\ln\frac{\lambda^2}{\lambda_0^2}
\label{eq:2.5} \\
\sigma(\mu_0,u) &=& -\frac{g^2}{4\pi^2}
\mu_0^2 \left( C_a (-\frac{u^2}{\mu_0^2}
+\ln\frac{u^2}{\mu_0^2}-\frac{5}{12})
+T_fN_f(\frac{1}{3}+\frac{m^2}{\mu_0^2}) \right)
\nonumber
\,,\end{eqnarray}
where scale $u$ has been introduced
\cite{ZhangHarindranath},
\cite{WilsonWalhoutHarindranathZhangPerryGlazek} 
to regulate 
small light-front $x$ divergences, 
$x\sim 0$ and $x\sim 1$, which correspond to
high light-front energies. 
In asymtotic free theories, such as QCD, the regulating scale 
can be related to the gauge invariant scale, 
using Callan-Semanzchik type equation.
Namely scale $u$ can be expressed through $\Lambda_{QCD}$,
solving the third order flow equations for 
the strong coupling constant $\alpha_s(\lambda)$.
Some calculations have been recently done in this direction
for the asymptotic free toy model \cite{Glazek2} and
for the three-gluon vertex in QCD \cite{Glazek3}.
However, here we do not perform these calculations 
and assume that the value $u$ is given by the hadron scale,
$u\sim \Lambda_{hadron}$.

The resulting effective Hamiltonian is defined
at the scale $\lambda\rightarrow 0$, 
therefore an effective gluon mass equals the 'physical' mass,  
$\mu^2_{ren}(\lambda=\lambda_0=0)=\mu_0^2$. In particular,
when the 'physical' mass is set to zero, $\mu_0=0$,
the effective QCD Hamiltonian has zero mass gauge fields, 
therefore our unitary transformaion does not violate gauge invariance
(though at the intermediate stages for finite $\lambda$
gauge invariance is broken by the cutoffs).
From Eq.~(\ref{eq:2.5}) one has in the limit $\mu_0\rightarrow 0$
\begin{eqnarray}
\sigma^{\prime}=\lim_{\mu_0\to 0}\sigma(\mu_0,u)=u^2\frac{g^2C_a}{2\pi^2}
\,,\label{eq:2.6}\end{eqnarray}
and, as shown below, 
$\sigma^{\prime}$ plays the role of the string tension 
between quark and antiquark. 
In Eq.~(\ref{eq:2.5}) the squared of the light-front cutoff, $u^2$,
defines the rate how fast the effective gluon mass
approaches asymtotically $\lambda\rightarrow\lambda_0\sim 0$
(from above $\lambda\geq\lambda_0$)
the 'physical' value $\mu_0$.
Asymptotic behavior of the effective gluon mass 
near the renormalization point 
$\lambda\rightarrow \lambda_0$ 
Eq.~(\ref{eq:2.5})
is important to take into account when
solving flow equations for effective quark interactions 
at vanishing gluon momenta.
In the next section this dependence, Eq.~(\ref{eq:2.5}), 
is used to find an effective 
quark potential, generated by flow equations.

\subsection{Effective quark-antiquark interaction}

Eliminating the quark-gluon coupling generates an effective interaction 
in the quark-antiquark sector. 
In the light-front frame an effective quark-antiquark 
interaction is given
\begin{eqnarray}
V_{q\bar{q}}=-4\pi\alpha_s C_f \langle\gamma^{\mu}\gamma^{\nu}\rangle
\lim_{(\mu_0,\lambda_0)\rightarrow 0} B_{\mu\nu}
\,,\label{eq:2.7}\end{eqnarray}
which includes 
dynamical interactions generated by flow equations 
and the instananeous term
present in the original light-front gauge  Hamiltonian. 
The gluon renormalization mass parameter 
('physical' gluon mass) $\mu_0$ and the renormalization point
$\lambda_0$ are set to zero at the end of calculations.
Here the current-current term in the exchange channel
reads
\begin{eqnarray}
\langle\gamma^{\mu}\gamma^{\nu}\rangle
=\frac{\left( \bar{u}(-k_{\perp},(1-x))\gamma^{\mu}
u(k_{\perp},x)\right)
\left(\bar{v}(k'_{\perp},x')\gamma^{\nu}v(-k'_{\perp},(1-x'))\right)}
{\sqrt{xx'(1-x)(1-x')}}
\,,\label{eq:2.8}\end{eqnarray}
where helicities of quarks are suppressed for simplicity;
and the interaction kernel is given
in full analogy with an effective electron-positron interaction
in the light-front frame \cite{Gubankova},
except for keeping the cutoff dependence in the four-momentum transfers,
as
\begin{eqnarray}
B_{\mu\nu}=g_{\mu\nu}\left(I_1+I_2\right)
+\eta_{\mu}\eta_{\nu} \frac{\delta Q^2}{q^{+2}}
\left(I_1-I_2\right)
\,,\label{eq:2.9}\end{eqnarray}
where $g_{\mu\nu}$ is the light-front metric tensor, 
and the light-front unity vector $\eta_{\mu}$
is defined as $\eta\cdot k=k^+$.
The cutoff dependence of four-momentum transfers along 
quark and antiquark lines is accumulated in the factor 
\begin{eqnarray}
I_1=\int_0^{\infty}d\lambda\frac{1}{Q_1^2(\lambda)}
\frac{df(Q_1^2(\lambda);\lambda)}{d\lambda}
f(Q_2^2(\lambda);\lambda)
\,,\label{eq:2.10}\end{eqnarray}
$I_2$ is obtained by the interchange of indices $1$ and $2$;
$f(z)$ is a similarity function;
and the light-front four-momentum transfers are defined 
\begin{eqnarray}
Q_1^2(\lambda) &=& Q_1^2+\mu_{ren}^2(\lambda)
\nonumber\\
Q_2^2(\lambda) &=& Q_2^2+\mu_{ren}^2(\lambda)   
\,,\label{eq:2.11}\end{eqnarray}
with 
\begin{eqnarray}
Q_1^2 &=& \frac{(x'k_{\perp}-xk'_{\perp})^2+m^2(x-x')^2}{xx'}
\nonumber\\
Q_2^2 &=& Q_1^2|_{x\rightarrow(1-x);x'\rightarrow(1-x')}
\,,\label{eq:2.12}\end{eqnarray}
where $\mu_{ren}$ is given in 
Eq.~(\ref{eq:2.5}); 
and
the average momenta are
$Q^2=(Q_1^2+Q_2^2)/2$ and $\delta Q^2=(Q_1^2-Q_2^2)/2$.
Calculating an effective kernel with explicit similarity functions,
one has (for details see \cite{GubankovaJiCotanch})
\begin{eqnarray}
\lim_{(\mu_0,\lambda_0)\rightarrow 0}B_{\mu\nu}
=g_{\mu\nu}\left(\frac{1}{Q^2}+\frac{\sigma'}{Q^4}\right)
+O\left((\frac{\delta Q^2}{Q^2})^n\right)
\,,\label{eq:2.13}\end{eqnarray}
where $n=2$ for smooth and $n=1$ for sharp cut-off functions,
and the first term does not depend 
on the cut-off function.
The four-momenta can be represented in the 'mixed' light-front
$(x,k_{\perp})$
and instant $\vec{k}=(k_{z},k_{\perp})$ 
frames as
\begin{eqnarray}
Q^2 &=& \vec{q}^{~2}-(2x-1)(2x'-1)(M_1-M_2)^2/4
\nonumber\\
\delta Q^2 &=& -(x-x')(M_1^2-M_2^2)/2
\,,\label{eq:2.14}\end{eqnarray}
where $\vec{q}=\vec{k}-\vec{k}^{\prime}=(q_z,q_{\perp})$
is the gluon three-momentum transfer,
and $M_1^2=4(\vec{k}^2+m^2)$ and $M_2^2=4(\vec{k}^{\prime 2}+m^2)$
are the total energies of the initial and final states.
Therefore in the limit of vanishing gluon momentum
$\vec{q}\rightarrow 0$, which defines mainly the bound state spectrum
because then the effective $q\bar{q}$ interaction is singular,
the four-momenta are $Q^2\rightarrow \vec{q}^{~ 2}$
and $\delta Q^2\rightarrow 0$, and the effective interaction
is given
\begin{eqnarray}
V_{q\bar{q}}=-\langle\gamma^{\mu}\gamma_{\mu}\rangle
\left(C_f\alpha_s\frac{4\pi}{\vec{q}^{~ 2}}+
\sigma\frac{8\pi}{\vec{q}^{~ 4}}\right)
\,,\label{eq:2.15}\end{eqnarray}
where a new constant $\sigma$ is introduced instead
of $\sigma'$, given by Eq.~(\ref{eq:2.6}),
as $\sigma=\sigma'\alpha_sC_f/2$.
One recovers the standard Coulomb and linear rising confining
potentials, $-C_f\alpha_s/r+\sigma\cdot r$, in configuration space.
It is remarkable, that, though calculations are done 
in the light-front frame, 
the result for the leading quark-antiquark 
effective interaction, Eq.~(\ref{eq:2.15}),
is rotationally invariant.
Confining term in Eq.~(\ref{eq:2.15}) 
with singular behavior like $1/\vec{q}^{~4}$,
arises from elimination of the quark-gluon coupling
at small gluon momenta, that is governed by the asymtotic behavior 
of the effective gluon mass.   

As was shown in \cite{Gubankova} with $\sigma=0$,
solving QED effective interaction for 
positronium spectrum numerically,
rotational symmetry holds with high accuracy for smooth 
cut-off functions and even for a sharp cutoff if the collinear
singular part is subtracted.
Based on our analyses here, 
we anticipate to see in numerical calculations, 
that also in the QCD effective interaction
rotationally nonsymmetric part contributes negligible
and meson spectrum manifests rotational invariance.

\section{Conclusions and outlook}

An effective QCD Hamiltonian in the light-front gauge
has been obtained, solving flow equations for the two lowest
Fock sectors selfconsistently. It has been shown that
it is possible to eliminate the minimal quark-gluon interaction
by using continuous unitary transformation. This elimination
causes the renormalization of the coupling functions of 
the Hamiltonian described by the flow equations. 
In the two lowest Fock sectors this change 
of the couplings corresponds to the renormalization
of the one-particle energies and to the generation 
of effective interactions between quarks, in particular
quark-antiquark interaction. In oder to set up 
these differential equations the generated new interactions, 
with more than three intermediate states, have to be neglected.
Truncating in number of particles participating in intermediate
states is not equivalent to perturbation theory 
in coupling constant, but rather is close to
Tamm-Dancoff approach.

Our approach has the following advantages:
(i) The original gauge field coupling can be completely
eliminated, even when the states connected by this
interaction are degenerate. The continuous transformation 
is chosen in such a way that the transformed Hamiltonian does not
contain any interations between one (anti)quark and the creation 
or annihilation of one gluon. These interactions,
connecting the states with energy differences less than
a cutoff scale $|E_p-E_q|\leq\lambda$, are still present
in similarity approach due to nelecting the renormalization
of single particle energies. They may cause mixing between
the low and high Fock sectors, and it is problematic to include them  
perturbatively. Ignoring these low-energy interactions 
may break the gauge invariance, and in the light-front frame  
the rotational symmetry.
(ii) Effective quark-antiquark interaction 
is rotationally symmetric at small gluon momenta $q$, 
while all collinear singular terms $\sim 1/q^+$ 
and $\sim 1/q^{+\, 2}$ cancel;
and contains in addition to a perturbative term $1/q^2$,
which can be obtained in the second order perturbation theory,
also a more singular behavior of $1/q^4$ type.  
Our result for the induced $q\bar{q}$-interaction differes also 
from the result of similarity scheme, where the collinear singular
part of the uncanceled instantaneous interaction
$\sim 1/q^{+\, 2}$, produces a logarithmic potential,
which is not rotationally symmetric.
The origin of these differences lies in the fact that in our approach
all couplings depend on $l$. In order to obtain properties 
(i) and (ii) the influence of the renormalization of the one-particle 
energies, in particular the gluon effective energy,
on the elimination of the quark-gluon coupling has to be taken
into account. By doing so the renormalization of the light-front 
gluon mass $\mu(\lambda)$ at zero gluon momentum 
in the asymptotic regime of small cutoff scales
is described by an integro-differential equation.
This equation can be solved, assuming the renormalization condition 
that the renormalized gluon mass is given 
at some small cutoff scale $\lambda_0$ by the 'physical' value $\mu_0$,
$\mu_{ren}(\lambda=\lambda_0)=\mu_0$. Renormalization
is understood in perturbative sense, i.e. the renormalized
effective gluon mass is obtained by absorbing the leading cutoff
dependence into the second order mass counterterm.
As a result the asymptotic behavior of the renormalized gluon mass
$\mu_{ren}(\lambda)$ for small cutoffs $\lambda$, 
approaching the renormalization point from above $\lambda\geq\lambda_0$,
has been obtained. As a consequence of the properties 
of $\mu_{ren}(\lambda)$ the quark-gluon coupling is always
eliminated even in the case of degeneracies, namely for vanishing
gluon momenta. A similar argumentation was used by Kehrein,
Mielke and Neu \cite{KehreinMielkeNeu} for the spin-boson
model, where the authors argued that the coupling to the bosonic bath
always is eliminated because of the renormalization of the tunneling
frequency. Also, in a complete analogous to our problem of interacting
electrons in BCS-theory, Lenz and Wegner \cite{LenzWegner} showed 
that the elimination of electron-phonon coupling 
for all states is a direct consequence 
of the renormalization of phonon frequency. 

Furthermore it is shown that due to the asymptotic behavior 
of the gluon energy, elimination of quark-gluon coupling at small 
gluon momenta gives rise to the enhancement of 'zero modes'
in the effective quark-antiquark interaction,
i.e. in the infrared region $q\sim 0$ a more singular potential
$1/q^4$, than in the perturbative case, is induced.
By discussing the consequences of this asymptotic behavior
it becomes clear that the approach of flow equations 
is superior to perturbative calculations, and probably also
to (perturbative) similarity scheme which works  
in terms of bare unrenormalized fields.
It can be seen that 
the flow of the coefficients (couplings) changes the generator $\eta$   
of the unitary transformation. Even if the flow of couplings 
is obtained within the perturbative frame, the unitary transformation
based on this generator $exp(\int dl \eta(l))$, 
includes infinite many orders
in perturbation theory, corresponding to (leading log)
resummation of diagrams.  
It is worth mentioning, that Lenz
and Wegner found in the system of 
interacting electrons \cite{LenzWegner}, 
that carrying out the $l$-ordering of $\eta$, the induced 
electron-electron interactions differ from the Fr{\"o}hlich's,
where the unitary transformation based on the second order
bound state perturbation theory is used. Moreover, this interaction 
is attractive in all momentum space, providing binding
electrons in Cooper pairs. Kehrein and Mielke obtained
similar modifications due to $l$-dependent generator 
by eliminating the hybridization term in the single impurity
Anderson model by continuous unitary transformations \cite{KehreinMielke}.
The authors showed that their approach generates a spin-spin
interaction which differs from the one obtained by the famous
Schrieffer-Wolff transformation. Their induced interaction
has the right high-energy cutoff, as compared to 
the Schrieffer-Wolff's result. Summarizing, 
within flow equations approach it is possible 
to make statements which can not be obtained by perturbation
theory.

Besides to this comparison with perturbative schemes, 
due to complete elimination of the quark-gluon coupling,
flow equation for an effective quark-antiquark interaction
can be integrated for all cutoffs down to $\lambda=0$.
In similarity approach one removes couplings perturbatively  
untill the finite cutoff, below which perturbation theory
breaks down. The choice of this cutoff depends on the 
problem considered, that might be ambiguous. 
In QCD treatment this cutoff 
introduces the nonzero scale in the theory, which breaks gauge 
and rotational invariance \cite{BrisudovaPerry}. 
In our approach, the regulator of small light-front $x$
sets up a scale in the effective theory, in particular
for the string tension in the effective quark-antiquark potential.
Besides of the nonzero scale, the resulting renormalized gluon mass
vanishes asymptotically, maitaining gauge invariance.
As a consequence, the effective quark-antiquark interaction
is rotationally symmetric.
However, in this work the small light-front $x$ cutoff scale $u$
enters as an input parameter, and is fitted to the string tension
from lattice calculations. To improve this, one should be looking 
for the fixed point of renormalization group and possible
relate the cutoff $u$ with renormalization group invariant
scale $\Lambda_{QCD}$. In this way one should include 
higher Fock sectors in the internediate states.
By doing so it is desirable to establish the independence
on the regularization scheme, used to regulate
small light-front momenta $x$.

The ultimate goal of the study is to solve the coupled chain
of flow equations in different sectors selfconsistently.
As has been shown, even an approximate solution
of the gluon gap equation together with the flow equation
for the effective interaction between probe quarks
provides some information beyond the perturbation theory.
The next step is to introduce dynamical quark degrees of freedom,
formulating quark gap equation, and study the influence
of the renormalization of the light-front quark mass
on the effective interaction between quarks.

It seems, that it is possible to isolate in the light-front frame 
the degrees of freedom which are responsible for the long-range
properties of QCD, and obtain some insight into the nonperturbative
QCD phenomena. 
This suggests that probably the light-front formulation may be 
the most suitable frame to solve the system of 
flow equations selfconsistently on computer.

\section*{Acknowledgments}
The author would like to thank the organizers of the workshop
for hospitality and support. The author is thanful to Stan Brodsky,
Stan Glazek, Igor Klebanov, Lev Lipatov and Zvi Bern
for helpful discussions. 
This work was supported 
by DOE grants DE-FG02-96ER40944, DE-FG02-97ER41048 and DE-FG02-96ER40947.

\end{document}